\documentclass[twocolumn,showpacs,aps,prl,epsfig,amssymb,amsmath,ulem,superscriptaddress]{revtex4-1}

\usepackage{color}
\usepackage{ifthen}     
\newboolean{SubmittedVersion}
\setboolean{SubmittedVersion}{False}    

\usepackage{graphicx}
\usepackage{amsmath}
\usepackage{amssymb}
\DeclareMathAlphabet{\mathpzc}{OT1}{pzc}{m}{it}

\def\Rb87{^{87}\text{Rb}}                     
\def\Na23{^{23}\text{Na}}                     
\def\Li6{^{6}\text{Li}}                       

{\catcode`\|=\active
  \gdef\Braket#1{\left<\mathcode`\|"8000\let|\BraVert {#1}\right>}}
\def\BraVert{\egroup\,\mid@vertical\,\bgroup}

\begin{document}

\title{Spin-orbit coupled Bose-Einstein condensates in a one-dimensional optical lattice}
\author{C.~Hamner}
\affiliation{Department of Physics and Astronomy, Washington State University, Pullman,  WA 99164, USA}
\author{Yongping~Zhang}
\affiliation{The University of Queensland, School of Mathematics and Physics, St Lucia, Queensland 4072, Australia}
\affiliation{Quantum Systems Unit, Okinawa Institute of Science and Technology, Okinawa 904-0495, Japan}
\author{M. A. Khamehchi}
\affiliation{Department of Physics and Astronomy, Washington State University, Pullman,  WA 99164, USA}
\author{Matthew J.~Davis}
\affiliation{The University of Queensland, School of Mathematics and Physics, St Lucia, Queensland 4072, Australia}
\author{P.~Engels}
\affiliation{Department of Physics and Astronomy, Washington State University, Pullman,  WA 99164, USA}
\email{engels@wsu.edu}

\begin{abstract}

The realization of artificial gauge fields and spin-orbit coupling for ultra-cold quantum gases promises new insight into paradigm solid state systems. Here we experimentally probe the dispersion relation of  a spin-orbit coupled Bose-Einstein condensate loaded into a translating optical lattice by observing its dynamical stability, and develop an effective band structure that provides a theoretical understanding of the locations of the band edges.  This system presents exciting new opportunities for engineering condensed-matter analogs using the flexible toolbox of ultra-cold quantum gases.
 
\end{abstract}
\pacs{03.75.Kk, 03.75.Mn, 03.75.Lm}
\maketitle

Spin-orbit coupling --- the interaction between a particle's spin and its mechanical motion --- plays a prominent role in  condensed matter physics \cite{Nagaosa}. Even though the spin-orbit  interaction is usually relatively weak, it can be important for bands close to the Fermi level \cite{Damascelli}. The combination of spin-orbit coupling with a periodic potential resulted in the prediction and discovery of topological insulators \cite{Kane1,Kane2}. Such spin-orbit coupled lattice systems, with the addition of strongly correlated many-body effects, can exhibit novel phases \cite{Balents}. These systems have transformed our understanding and classification of insulators and have become a significant focus of recent research \cite{Hasan,Shoucheng}. They afford the possibility of studying new phase transitions and realizing exotic spin models \cite{Jackeli}.

Simulating model Hamiltonians relevant to condensed matter physics has developed into a major area of research for experiments with dilute gas Bose-Einstein condensates and degenerate Fermi gases ~\cite{Lewensteinreview,Blochreview}.  Quantum gases in optical lattices are nearly disorder free and often exhibit long coherence times~\cite{Lewensteinreview,Blochreview}. Additionally, quantum gases allow the modification of the interparticle interactions, e.g. by tuning the two-body scattering length \cite{Collapse} or engineering long range dipolar interactions \cite{DipolarBECStability}, creating great flexibility for implementing model Hamiltonians.  While many electronic condensed-matter systems naturally exhibit a band structure due the periodicity of an underlying crystal lattice, in ultra-cold quantum gases band structures can be engineered by loading the system into an optical lattice. Both an optical lattice potential~\cite{Inguscio} and spin-orbit coupling~\cite{StamperKurnHigbie} can strongly modify the single-particle dispersion relation of a quantum gas, resulting in novel band structures.

\begin{figure}
\begin{center}
\ifthenelse{\boolean{SubmittedVersion}}{}{\includegraphics[width=1.0\linewidth]{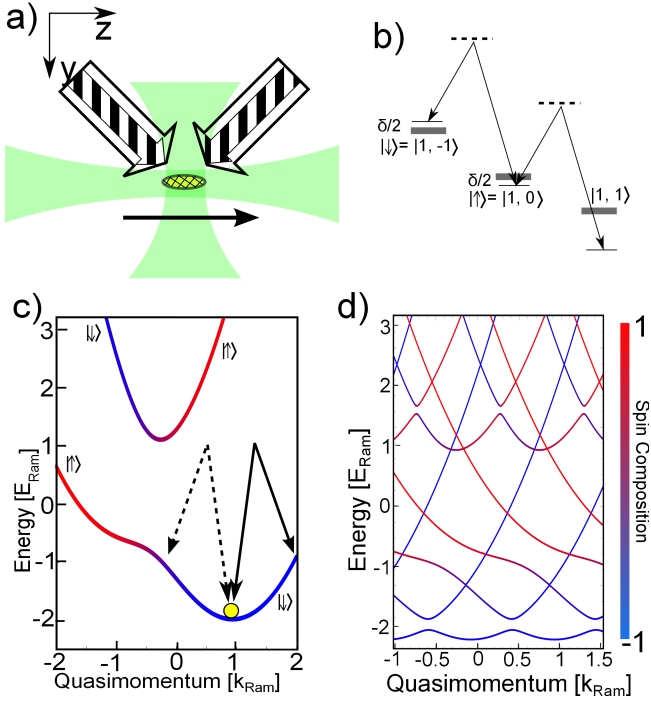}}
\end{center}
\caption{(Color online) Spin-orbit coupled $^{87}$Rb BEC in a one-dimensional optical lattice. (a) Experimental geometry. The BEC (yellow hashed) is confined in an optical dipole trap (solid green). Two sets of laser beams intersect the BEC at a 45$^\circ$ angle,  generating the spin-orbit coupling (white arrow) and a translating optical lattice (striped arrows). (b)  Raman coupling scheme in the F=1 manifold of $^{87}$Rb with detuning $\delta$. (c) Typical band structure $E_{\pm}(k_z)$ of $H_{\rm SOC}$ with the color (grey scale) indicating the spin-polarization, defined as the relative population difference of the bare spin components $(|\psi_{\uparrow}|^2-|\psi_{\downarrow}|^2)/(|\psi_{\uparrow}|^2+|\psi_{\downarrow}|^2)$. The BEC is prepared at the minimum of lower band (circle). The arrows indicate a possible two photon coupling due to the lattice translating with negative (dashed) and positive (solid) velocity.  (d) Bloch spectrum of the stationary optical lattice in the presence of spin-orbit coupling. The lines correspond to $E_{\pm}(k_z)$ and $E_{\pm}(k_z+2 n k_{\rm lat})$, where $n$ is an integer.  The spin composition is encoded in the line color (grey scale). The parameters used for (c) and (d) are $\hbar \delta=1.6~E_{\rm Ram}$, $\hbar \Omega=2~E_{\rm Ram}$ with the additional parameters $U_0=-1.4E_{\rm lat}$ and $v=0$ for (d).}
\label{setup}
\end{figure}

\begin{figure}
\includegraphics[width=1.0\columnwidth]{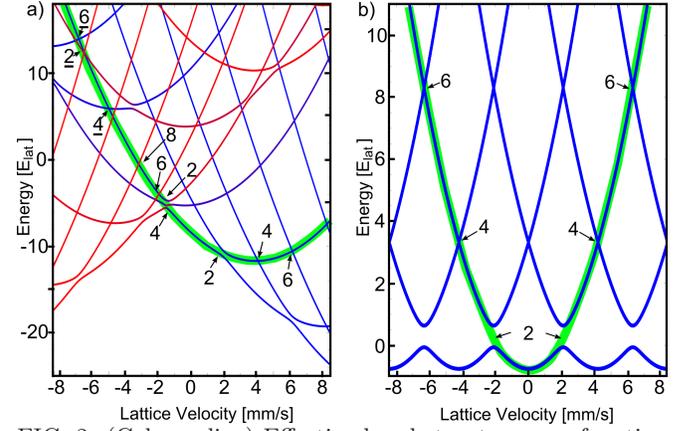}
\caption{(Color online) Effective band structure as a function of the lattice velocity. The thick green (grey) lines indicate the position at which the BEC is placed in the experiments.  (a) BEC with spin-orbit coupling and $\hbar\delta=1.6E_{\rm Ram}$ as shown in Fig.~\ref{DI}(b). (b) BEC without spin-orbit coupling as in Fig.~\ref{DI1}.  The numbers in the graphs indicate the order of the associated multi-photon resonances.}
\label{Trajectories}
\end{figure}

In this letter we perform a detailed study of a Bose-Einstein condensate (BEC) with spin-orbit coupling \cite{StamperKurnHigbie,Lin,Jing,Shuai,Peter} loaded into a shallow, translating one-dimensional optical lattice. We find that the system exhibits a number of dynamical instabilities induced by the periodic dispersion relation of the lattice \cite{Zhang}.  The instabilities are marked by an initial exponential growth of excitations in the BEC, and are most significant in the vicinity of a band gap.  We characterize the strengths of the instabilities by the loss rate of condensate atoms and find that a dynamical instability is present for lattice velocities exceeding a critical velocity within the first Brillouin zone \cite{Biao, Inguscio}.  The strength of the instability depends on both the lattice speed and direction of motion.  This is an indicator of the lack of Galilean invariance in the presence of the spin-orbit coupling \cite{Qizhong}. We compare our results with a Bogoliubov analysis of the system, finding good agreement.

We begin by providing a brief description of our experimental setup --- full details can be found in the Supplementary Material~\cite{Suppmatt}.  Spin-orbit coupling in BECs can be induced by Raman dressing schemes~\cite{Lin,Dalibard,Galitski}, and the geometry of our experiment is shown schematically in Fig.~1(a).  The Raman lasers couple the $| 1, -1 \rangle =  |\!\!\downarrow\rangle$ and $| 1, 0 \rangle = |\!\!\uparrow\rangle$ states of a $^{87}$Rb BEC in the F=1 hyperfine manifold.  A 10 G bias magnetic field causes a sufficiently large quadratic Zeeman splitting such that the $|1,+1\rangle$ state is far from resonance and can be eliminated, hence realizing an effective spin-$1/2$ system. The system without the one-dimensional lattice is modeled by the single-particle  Hamiltonian $H_{\rm SOC}=\frac{\hbar^2 k^2_z}{2m}+\gamma p_z\sigma_z+\frac{\hbar\delta}{2}\sigma_z+\frac{\hbar \Omega}{2}\sigma_x$ \cite{Lin}.  Here $m$ is the atomic mass, $\hbar k_z$ is the quasimomentum in the spin-orbit direction and $\{\sigma_i\}$ are the Pauli matrices. The spin-orbit coupling strength is $\gamma=\hbar k_{\rm Ram}/m$, where $k_{\rm Ram}$ is the wavevector of the Raman beams in the $z$-direction, $\delta$ is the detuning, and $\Omega$ is the Rabi frequency.  A typical band structure for our parameters is shown in Fig.~\ref{setup}(c), where the band energies are  $E_{\pm}(k_z)=\frac{\hbar^2k_z^2}{2m} \pm \hbar \sqrt{(\gamma k_z+\frac{\delta}{2})^2+\frac{\Omega^2}{4}}$. 

\begin{figure*}
\begin{center}
\includegraphics[width=0.99\columnwidth]{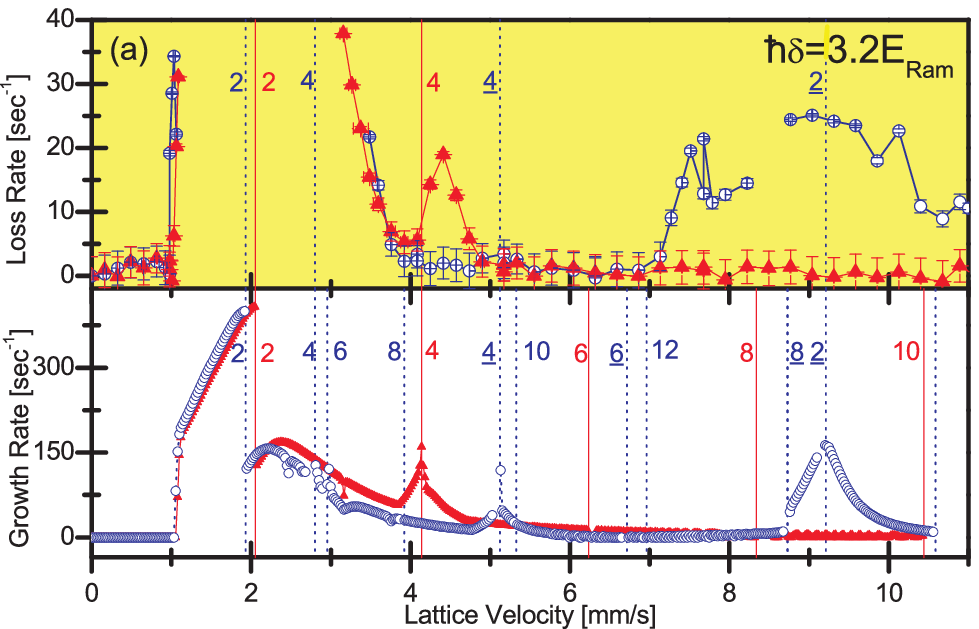}
\includegraphics[width=0.99\columnwidth]{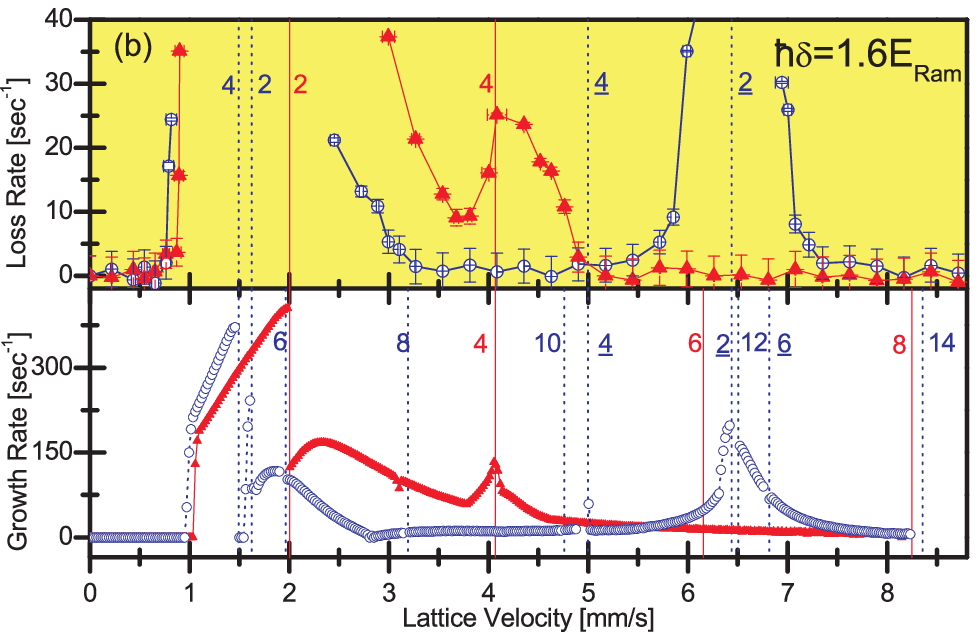}
\includegraphics[width=0.99\columnwidth]{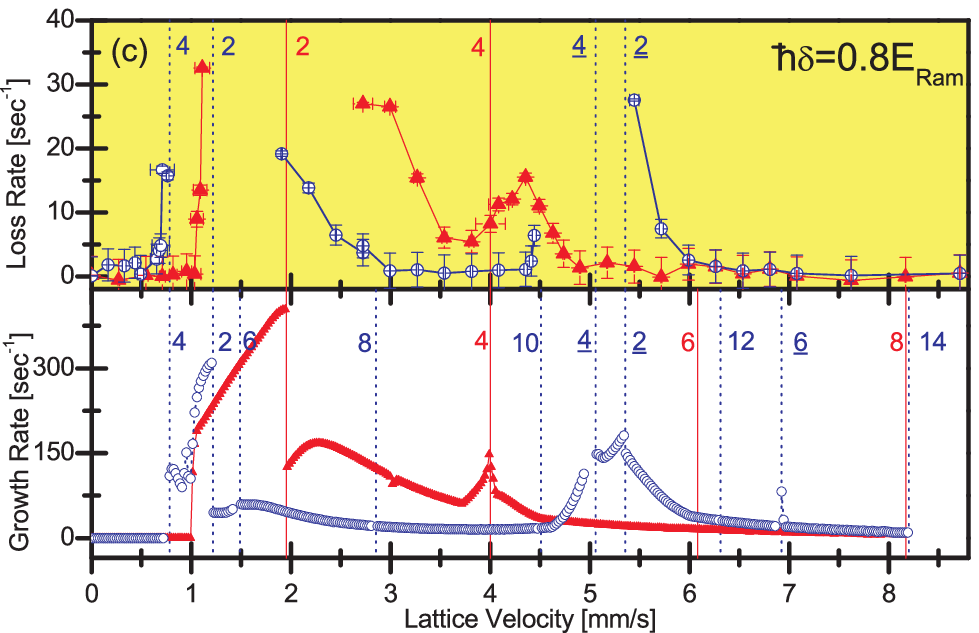}
\includegraphics[width=0.99\columnwidth]{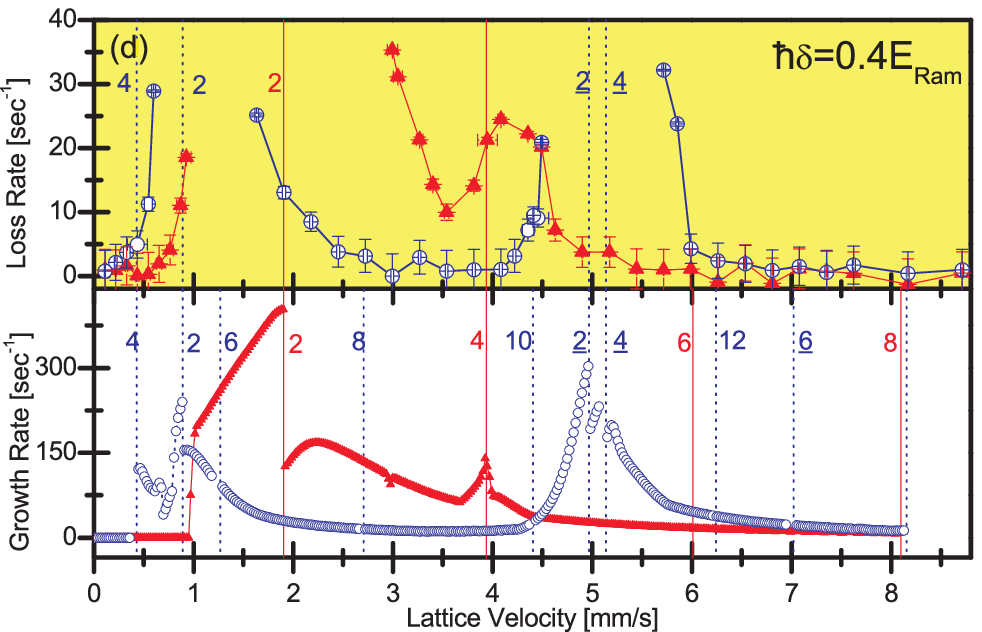}
\end{center}
\caption{(Color online) Dynamical instability of the spin-orbit coupled BEC as a function of lattice speed with (a--d) $\hbar\delta/E_{\rm Ram} = \{3.2, 1.6, 0.8, 0.4\}$ respectively. The strength of the dynamical instability is measured experimentally by the loss rate of atoms in the BEC (upper panels), while theoretically it is represented by the largest growth rate of Bogoliubov excitations (lower panels). Each resonance (vertical line) is labeled with the number of photons generating the band edge, with underlined integers denoting resonances between the upper and lower spin-orbit bands.  The solid red triangles (open blue circles) indicate the positive (negative) direction of the lattice motion.} \label{DI}
\end{figure*}

Two additional laser beams with $\lambda_{\rm lat}\approx 1540$ nm and small frequency difference $\Delta \nu$  generate the translating optical lattice. The lattice beams are collinear with the Raman lasers such that $k_{\rm lat}=2\pi/(\lambda_{\rm lat}\sqrt{2})$.  The single-particle Hamiltonian of the spin-orbit coupled lattice system is  $H_{\rm sp}=H_{\rm SOC}+U_0\sin^2[k_{\rm lat}(z-vt)]$. The lattice velocity  $v=\pi  \Delta \nu/k_{\rm lat}$ can be adjusted by varying the frequency difference $\Delta \nu$ between the two lattice beams.  For the experiments presented in this manuscript, $U_0=-1.4E_{\rm lat}$, where $E_{\rm lat}=\frac{\hbar^2 k_{\rm lat}^2}{2m}$. The presence of the optical lattice extends the spin-orbit coupled bands in Fig.~\ref{setup}(c) to the Bloch spectrum in Fig.~\ref{setup}(d). In the repeated zone scheme the Bloch spectrum is constructed through copies of the spin-orbit bands shifted by integers of the reciprocal lattice vector $2 n k_{\rm lat}$ in quasimomentum and $2 n \hbar k_{\rm lat} v$ in energy, where $n=0,\pm1, \pm2,\ldots$. Gaps  open in the Bloch spectrum wherever intersections between $E_{\pm}(k_z )$ and $E_{\pm}(k_z + 2nk_{\rm lat})-2n \hbar k_{lat} v$ occur. The width of the gap depends on the lattice depth $U_0$ and the overlap between the spin composition of the states coupled by the lattice beams.   Typically the gap width corresponding to $|n|$ is larger than that corresponding to $|n|+1$.  This is evident in Fig.~\ref{setup}(d), where the energy gaps are largest  for $|n|=1$ in both the lower as well as the upper dressed bands. Physically, the band gaps can be understood from multi-photon resonances in which the momentum of the atoms can be changed coherently by multiples of the reciprocal lattice vector $2 n \hbar k_{\rm lat}$.  

Before describing the experimental results, it is instructive to introduce an effective band structure picture. As the translating optical lattice potential is time dependent in the lab frame, it is convenient to go into the frame in which the optical lattice is stationary.  This results in the Hamiltonian $H^M_{\rm sp}=H_{\rm SOC}-vp_z+U_0\sin^2(k_{\rm lat} z)$.  With a simple Galilean substitution, $P=p_z-mv$, one obtains $\bar{H}^M_{\rm sp}=\frac{P^2}{2m}+\gamma P\sigma_z+(\delta+2 m \gamma v / \hbar )\frac{\hbar}{2}\sigma_z+\frac{\hbar \Omega}{2}\sigma_x+U_0\sin^2(k_{\rm lat}z)$ (where we have left out a constant energy term $mv^2/2$.) In addition to the lattice potential, $\bar{H}^M_{\rm sp}$ is non-trivially different from $H_{\rm SOC}$ as the term $\delta + 2 m \gamma v /\hbar$, which can be interpreted as an effective detuning of the Raman beams, is dependent on the frame velocity $v$.  This is due to the broken Galilean invariance of the spin-orbit coupled BEC.  Physically this arises because the Raman lasers generating the spin-orbit coupling provide a fixed frame of reference.  

To understand the band structure, we trace the location of the BEC in the single-particle band spectra as a function of the lattice velocity. The BEC is initially assumed to be in the ground state of the spin-orbit coupled band $E_-(k_{z})$ with a finite quasimomentum, $k_{\rm min}$, which is approximately conserved when the optical lattice is introduced~\cite{Biraben}.  The energies $E^M (k_{\rm min},v)$, taken from the Bloch spectrum of $H_{\rm sp}^M$ at $k_{\rm min}$, are shown as a function of lattice velocity in Fig.~\ref{Trajectories}(a) for $\hbar\delta = 1.6~E_{\rm Ram}$, $\hbar\Omega = 2~E_{\rm Ram}$, and $U_0 = -1.4E_{\rm lat}$, where $E_{\rm Ram} = (\hbar k_{\rm Ram})^2/2m$.  We label the avoided crossings within these effective band structures by integers $2n$ that indicate the photon processes involved, $E_-^M (k_{\rm min},v)=E_-^M (k_{\rm min}\pm 2n k_{\rm lat},v)$. Resonances occurring between the lower and upper spin-orbit bands [$E_-^M (k_{\rm min},v)=E_{+}^M (k_{\rm min}\pm 2n k_{\rm lat},v)$] are denoted by an underlined number $\underline{2n}$. When the spin-orbit coupled BEC is adiabatically loaded into the translating lattice, it occupies a state near the free particle dispersion (thick green line). It is interesting to note that the ordering of the band edges is not straightforward, and the positions of the band edges are not equally spaced.  The exact ordering and position strongly depend on the chosen parameters $\delta$, $\Omega$, and the ratio $k_{\rm lat}/k_{\rm Ram}$. For comparison, Fig.~\ref{Trajectories}(b) presents the analogous band structure for a BEC in a translating lattice but without spin-orbit coupling.  As is well-known in this case, the effective band structure and the BEC location (thick green line) are symmetric about the direction of motion, the band edges are equally spaced, and the effective dispersion relation recovers the Bloch spectrum.

\begin{figure}
\begin{center}
\includegraphics[width=0.99\columnwidth]{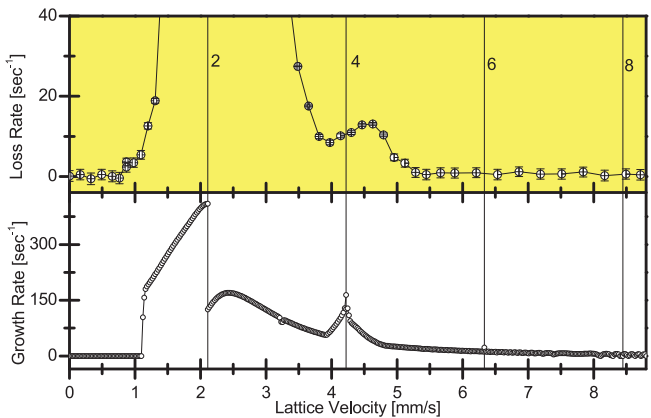}
\end{center}
\caption{Dynamical instability of the BEC without spin-orbit coupling as a function of lattice velocity.  The strength of the dynamical instability is measured experimentally by the loss rate of atoms in the BEC (upper panels), while theoretically it is represented by the largest growth rate of any Bogoliubov excitation (lower panels). Each resonance (vertical line) is labeled with the number of photons generating the band edge.} \label{DI1}
\end{figure}

To experimentally probe the dynamical instability we measure the loss spectra for the system as a function of the velocity of the optical lattice.  Our experiments begin with a nearly pure $^{87}$Rb BEC in the ground state in the presence of spin-orbit coupling with $\hbar \Omega=2~E_{\rm Ram}$  before ramping up a translating optical lattice with $U_0=-1.4E_{\rm lat}$~\cite{Suppmatt}.  We hold the BEC in the lattice potential for 100~ms, during which excitations caused by instabilities can grow and population is lost from the BEC.
The experimental results for the loss rate as a function of the lattice velocity, for four different values of the Raman detuning $\delta$, are plotted in the upper panels of Fig.~\ref{DI}(a--d).  In the absence of  Galilean invariance, we must differentiate between the two translating directions for the optical lattice.  In Fig.~\ref{DI} we plot negative (positive) velocities in dashed open blue circles (solid red triangles), corresponding to the dashed (solid) arrows in Fig.~\ref{setup}(c).  With this convention a lattice translating in the positive direction couples to states that resemble free particles, while a lattice translating in the negative direction couples to states that are strongly modified by the spin-orbit coupling. For comparison, we have also performed these measurements without spin-orbit coupling (see Fig.~\ref{DI1}) and for this case find agreement with prior experimental work \cite{Inguscio}. 

We model our experiment using the Bogoliubov-de Gennes (BdG) equations based on a one-dimensional mean-field description of a homogeneous BEC \cite{Suppmatt}. We identify the quasi-particle mode with the largest imaginary part of the energy, corresponding to the largest initial growth rate, and plot this rate as a function of velocity in the lower panels of Fig.~\ref{DI}(a--d). The theoretical results provide a good understanding of the experimental measurements.  While the theoretically calculated growth rates are not identical to the experimental loss rates presented in the upper panels of Fig.~\ref{DI}, they have previously been found to be a reasonable indication of the strength of dynamical instability~\cite{Inguscio}.    Both the experimental data and the numerical results demonstrate that the critical speed for the onset of the dynamical instability is different for the two directions of motions. This is particularly evident in the experimental and numerical results for the smaller detunings of $\hbar\delta=0.8~E_{\rm Ram}$ and $\hbar\delta=0.4~E_{\rm Ram}$ in Fig.~\ref{DI}(c,d) near $v=\pm0.5$ mm/s, where the critical velocity is smaller for the negative direction.  Above the critical velocity the dynamical instability is most significant in the vicinity of the band edges. Loss occurs in all higher bands as well, but the loss rate in higher bands is significantly reduced.

The dynamical stability of the BEC is quite different for the two directions of motion. In Fig.~\ref{DI}(a--d) the behavior of the loss and growth rates for the positive direction of motion (red solid triangles) is very similar to that of the case without spin-orbit coupling shown in Fig.~\ref{DI1}.  However, in the negative direction of motion (blue open circles) the behavior is strongly modified.  For example in Fig.~\ref{DI}(a) for $\hbar \delta = 3.2~E_{\rm Ram}$, a pronounced additional loss feature appears centered around $v=9$ mm/s, shifting to smaller velocities for smaller $\delta$ in Fig.~\ref{DI}(b--d).  This feature is caused by the two photon resonance $\underline{2}$ between $E_-^M (k_{\rm min},v)$ and $E_{+}^M (k_{\rm min}-2k_{ \rm lat},v)$ (i.e. the lattice resonance between the lower and upper spin-orbit bands).  For comparison, the large loss feature near $v=2$ mm/s is due to the $2$ photon resonance within the lowest spin-orbit band.  Even though both of these loss features arise from two photon couplings, the $\underline{2}$ feature is weaker. This is in part due to the reduced overlap of the spin composition between $ E_-^M (k_{\rm min},v)$ and $E_{+}^M (k_{\rm min}-2k_{\rm lat},v)$.  For the positive direction of motion of the lattice in Fig.~\ref{DI}(a--d) the $\underline{2}$ resonance between  $E_-^M (k_{\rm min},v)$ and $E_{+}^M (k_{\rm min}+2k_{\rm lat},v)$ occurring at large velocity is suppressed by the small overlap in spin composition for our chosen parameters.  For example, with $\hbar\delta=1.6$~$E _{\rm Ram}$, such a resonance occurs at $v=21.6$~mm/s but the modification to the Bloch spectrum is negligible.  Another loss feature near $v=4.5$ mm/s in Fig.~\ref{DI}(a) in the positive direction corresponds to the $4$ photon resonance, and is shifted to smaller velocities in the negative direction.  In the experimental results for the negative direction it cannot be differentiated from the dominant $2$ band edge, and is diminished due to the smaller overlap of the spin compositions.

In conclusion, we have studied the rich dispersion relation of a spin-orbit coupled BEC in a weak optical lattice by probing the losses of the system as a function of lattice velocity. Our study opens the door to investigations on the effect of spin-orbit coupling for the superfluid-to-Mott insulator transition~\cite{Cole} and the physics of strong spin-orbit correlated Mott insulators \cite{Radic}.  Our methods have a potential application for the realization of the fractional quantum Hall affect without Landau levels using a spin-orbit coupled lattice~\cite{FluxLattice}.\\

\noindent\textbf{Acknowledgements:} PE, MAK, and CH acknowledge funding from NSF and ARO.  YZ and MJD acknowledge financial support
from the Australian Research Council Discovery Project DP1094025.\\

\section{Supplementary Materials:} 

\section{Theoretical stability analysis} 

To model the experimental results, we perform a stability analysis in the mean-field treatment for the homogeneous BEC. We start from the  one-dimensional Gross-Pitaevskii equation (GPE) in dimensionless form
\begin{equation}
\label{GP} i\frac{\partial\Psi}{\partial t}=[H_{\rm
sp}^M+c(|\Psi_{\uparrow}|^2+|\Psi_{\downarrow}|^2)]\Psi,
\end{equation}
where $\Psi = [\Psi_{\uparrow}(z),\Psi_{\downarrow}(z)]^T$ with $H_{\rm sp}^M=-\frac{1}{2}\frac{\partial^2}{\partial z^2}-i\gamma\frac{\partial}{\partial z}\sigma_z+\frac{\delta}{2}\sigma_z +\frac{\Omega}{2}\sigma_x+iv\frac{\partial }{\partial z}+U_0\sin^2(z)$. The units of energy, time and length are $2~E_{\rm lat}$, $m/ \hbar k^2_{\rm lat}$ and $1/k_{\rm lat}$ respectively. The dimensionless quantities are defined as $\gamma=k_{\rm Ram}/k_{\rm lat}$ and $v=\pi m \Delta \nu / \hbar k^2_{\rm lat}$. The parameter $c$ is the nonlinear coefficient that is proportional to the s-wave scattering length. For this analysis we approximate the system by setting all of the s-wave scattering lengths equal. This is a reasonable assumption for the chosen experimental states where the differences in intra- and inter-component scattering lengths are small.

We calculate the Bloch spectrum and corresponding Bloch states of the GPE using the normalization condition within a unit cell. We choose a weak nonlinearity $c=0.05$ such that our nonlinear Bloch states converge for all lattice velocities. We have verified that our results are qualitatively independent of the exact value of $c$. The dynamical stability of the Bloch states is analyzed using linear perturbation theory and making the ansatz
\begin{equation}
\label{perturbation} \Psi=e^{-i\mu t +ikz}[\phi+Ue^{iqz-i\omega
t}+V^*e^{-iqz+i\omega^* t}].
\end{equation}
The Bloch state of interest is $\exp(-i\mu t +ikz) \phi$ where $\phi=(\phi_{\uparrow}, \phi_{\downarrow})^T$,  $\mu$ is the chemical potential and $k$ is the quasimomentum of the Bloch state into which the BEC is loaded in the experiment.  $U=(U_{\uparrow}, U_{ \downarrow})^T$ and $V=(V_{\uparrow}, V_{\downarrow})^T$ are perturbation amplitudes, and $q$ and $\omega$ are the quasimomentum and frequency of the perturbation respectively. Substituting Eq.~(\ref{perturbation}) into Eq.~(\ref{GP}) results in the Bogoliubov de Gennes (BdG) equations
\begin{equation}
\omega\begin{pmatrix} U_{\uparrow} \\  V_{\uparrow}  \\
U_{\downarrow}   \\  V_{\downarrow} \end{pmatrix}= \mathcal{H}
\begin{pmatrix} U_{\uparrow} \\  V_{\uparrow}  \\  U_{\downarrow}   \\  V_{\downarrow} \end{pmatrix},
\label{s3}
\end{equation}
with the BdG Hamiltonian
\begin{widetext}
\begin{equation}
\mathcal{H}=
\begin{pmatrix} \mathcal{H}_{\uparrow}(k,q) &c\phi_{\uparrow}^2&\frac{\Omega}{2}+c\phi_{\uparrow}\phi_{\downarrow}^*&c\phi_{\uparrow}\phi_{\downarrow} \\
                           -c\phi_{\uparrow}^{*2} & -\mathcal{H}_{\uparrow}^* (k,-q) & -c\phi_{\uparrow}^*\phi_{\downarrow}^* & -\frac{\Omega}{2}-c\phi_{\uparrow}^*\phi_{\downarrow}\\
                           \frac{\Omega}{2}+c\phi_{\uparrow}^*\phi_{\downarrow}& c\phi_{\uparrow}\phi_{\downarrow}& \mathcal{H}_{\downarrow}(k,q) & c\phi_{\downarrow}^2 \\
                           -c\phi_{\uparrow}^* \phi_{\downarrow}^* & -\frac{\Omega}{2}-c\phi_{\uparrow}\phi_{\downarrow}^*&-c\phi_{\downarrow}^{*2}& -\mathcal{H}_{\downarrow}^*(k,-q)
\end{pmatrix}.
\end{equation}
Here
\begin{align}
\mathcal{H}_{\uparrow}(k,q)&=\mathcal{H}_0(k,q)+\frac{\delta}{2}-i\gamma
[\frac{\partial}{\partial
z}+i(k+q)]
+2c|\phi_{\uparrow}|^2+c|\phi_{\downarrow}|^2,\\
\mathcal{H}_{\downarrow}(k,q)&=\mathcal{H}_0(k,q) - \frac{\delta}{2}+
i\gamma [\frac{\partial}{\partial
z}+i(k+q)]
+c|\phi_{\uparrow}|^2+2c|\phi_{\downarrow}|^2,
\end{align}
with 
\begin{equation}
\mathcal{H}_0(k,q)=-\frac{1}{2}     [\frac{\partial}{\partial
z}+i(k+q-v)]^2-\frac{v^2}{2}+U_0\sin^2(z)-\mu.
\end{equation}
\end{widetext}
The system of equations~(\ref{s3}) is then solved for the eigenvalues $\omega$.  If any $\omega$ is complex-valued, the Bloch state $\phi$ is dynamically unstable. The growth rate of excitations in the system is estimated as the largest imaginary value of any $\omega$, as this will dominate the growth.

\begin{figure*}[t]
\includegraphics[width=1.5\columnwidth]{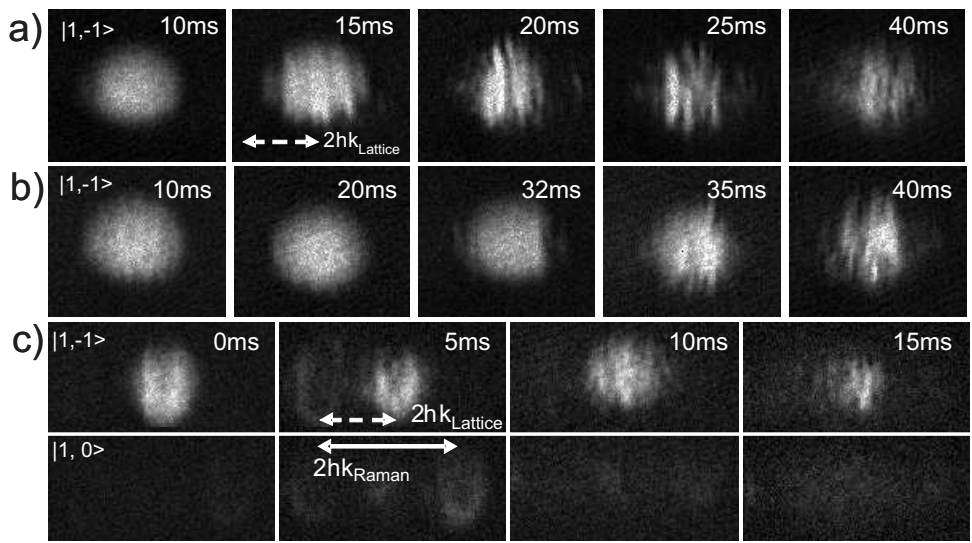}\\[2mm]
\caption{Absorption images revealing the density dynamics in the translating optical lattice. The hold time indicated  is measured from the end of the 10 ms loading process. (a) BEC without spin-orbit coupling, $v=1.3$~mm/s. (b) Spin-orbit coupled BEC with $\hbar\delta=1.6~E_{ \rm Ram}$ and $v=+1.1$~mm/s. (c) Spin-orbit coupled BEC with $\hbar\delta=1.6~E_{\rm Ram}$ and $v=-6.5$~mm/s, exhibiting spin dynamics. } \label{SpinDynamics}
\end{figure*}

To clearly convey the underlying physics, a single-particle description is used in the analysis in the main text.  However, for the stability analysis shown in Fig.~3 and Fig.~4 of the main text, the nonlinearity of the system is taken into account.  The use of the single particle description is justified as the nonlinearities due to repulsive interactions primarily contribute an overall shift to the Bloch spectrum~\cite{Stringari}.

\section{Experimental methods} 

Our experiment begins with a  BEC of $^{87}$Rb containing approximately $\sim1\times10^5$ atoms in the $|1, -1 \rangle$ hyperfine state. The BEC is held in an optical dipole potential with harmonic trapping frequencies $(\omega_x, \omega_y,\omega_z)=2\pi \times (134,170, 20)$ Hz, where $\hat{x}$ is the vertical direction. Two crossed Raman lasers with a wavelength in the range of $\lambda_{\rm Ram}=784-789$ nm are shone onto the BEC and propagate along the $\mathbf{e}_y\pm \mathbf{e}_z$ direction respectively with a relative angle $\pi/2$. The Raman lasers couple the $| 1, -1 \rangle =  |\!\!\downarrow\rangle$ and $| 1, 0 \rangle = |\!\!\uparrow\rangle$ states in the F=1 hyperfine manifold.  A 10 G magnetic bias field causes a quadratic Zeeman splitting of $7.4~E_{\rm Ram}$, where $E_{\rm Ram} = (\hbar k_{\rm Ram})^2/2m$, so that the $|1,+1\rangle$ state is far from resonance and can be eliminated. This realizes an effective spin-$1/2$ system. Two additional laser beams with $\lambda_{\rm lat} \approx 1540$ nm and small frequency difference $\Delta \nu$  generate the translating optical lattice. The lattice beams are collinear with the Raman lasers such that $k_{\rm lat}=2\pi/(\lambda_{\rm lat}\sqrt{2})$.  The ratio of the recoil momenta is $k_{\rm Ram}/k_{\rm lat}=1.96$, and the recoil energies are  $E_{\rm Ram}= h \times 1.86$~kHz  and $E_{\rm lat}=h \times 483$ Hz.  The spin-orbit coupled BEC is adiabatically loaded into the optical lattice by linear ramps of the lattice depth over 10 ms from $U_0=0$ to $U_0=-1.40 \pm 0.15~E_{\rm lat}$.  The adiabaticity is verified by turning on and off the lattice in two subsequent 2 ms long linear ramps of the intensity and checking for the absence of heating and additional momentum components.  Only velocities where no population transfer to additional momentum states and no heating of the BEC is  observed are utilized in these experiments. This check is conducted for BECs both with and without spin-orbit coupling. Following each experimental realization, the Raman and lattice beams are jumped off projecting the BEC onto the bare hyperfine and momentum states  followed by Stern-Gerlach separation during 11.5 ms time-of-flight period and absorption imaging.

\begin{figure*}[t]
\begin{center}
\includegraphics[width=1.8\columnwidth]{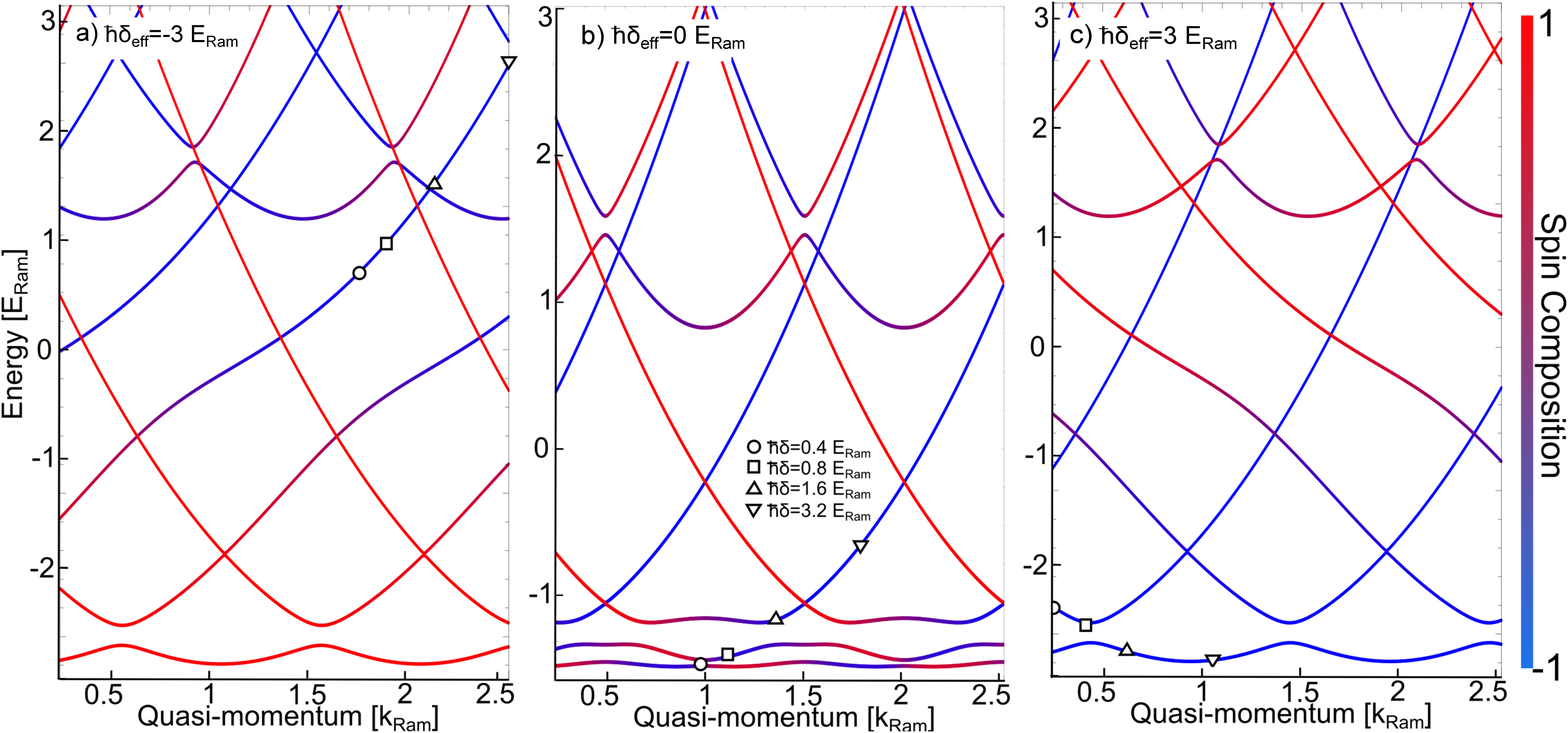}
\end{center}
\caption{(Color online) Band structure for three different values of the effective detuning $\delta_{\rm eff}$.  (a--c) are for $\hbar \delta_{\rm eff}/E_{\rm Ram}= \{-3,0,3\}$ respectively.  The calculated location of the BEC within these band structures is indicated for the values of $\delta$ presented in the main text.  Here, $\hbar\delta/E_{\rm Ram} = \{0.4, 0.8, 1.6,3.2\}$ are indicated by the circles, squares, triangles pointing up, and triangles pointing down respectively.  The corresponding lattice velocities are determined by $\delta_{\rm eff}=\delta + 2m \gamma v /\hbar$, and the spin composition is indicated by the colour of the band.} 
\label{SuppBandStructure}
\end{figure*}

\subsection{Further experimental considerations}

The analysis presented in the manuscript assumes that the minimum of the lower spin-orbit coupled band, $E_{-}(k_z)$, at $k_z=k_{\rm min}$ is occupied by the BEC. For the experimental realization, a nearly pure BEC (with minimal thermal component) is indeed loaded into the minimum of the spin-orbit band structure at $k_z=k_{\rm min}$. However, for some negative velocities taken at the smallest Raman detuning $\hbar\delta = 0.4 \pm 0.1 E_{\rm Ram}$ there is a small fraction of the atoms occupying a quasimomentum near $-k_{\rm min}$, following the 100 ms in the translating optical lattice.  For all other experimental data presented in the main text, the BEC remains near $k_{\rm min}$ for the duration of the experiment.  In all experimental images taken after the wait time in the translating optical lattice, a discernible thermal cloud is observed. These experimental considerations are not accounted for in the BdG analysis.

For numerous detunings and lattice velocities we have verified that the losses during the wait time in the translating optical lattice fit well to an exponential decay. The loss rates presented in Fig.~3 and Fig.~4 of the main text are determined by measuring the atom number after the 100 ms wait time for both the stationary lattice and for the translating lattice. Assuming  exponential decay, we calculate the contribution to the loss rate from the moving lattice.  These are the quantities plotted in Fig.~3 and Fig.~4. The vertical error bars reflect the uncertainty in the measurement of the atom numbers. Long term drifts in these measurements are suppressed by alternating the direction of subsequent optical lattice velocities during experimental realizations. The horizontal error bars reflect the uncertainty of the quasimomentum due to mismatch of the group velocities for the system with and without the lattice~\cite{Inguscio}. For the data presented in Fig.~3 we only probe velocities where the momentum imparted by the translating lattice is small.

The discussion in the manuscript focuses on the role of the dynamical instability.  However, we note that an energetic instability is also present in this system as well, but that it does not play a significant role for the observed dynamics.  The energetic instability is suppressed in our experiments by the small initial thermal fraction and the slower timescales associated with atom losses caused by the energetic instability~\cite{Inguscio}.  The dominant nature of the dynamical instability is made evident by the observation of density modulations in the BEC \cite{Inguscio} as shown in Fig.~\ref{SpinDynamics}. In Fig.~\ref{SpinDynamics}(a) we present an example  for the case of a BEC in a translating optical lattice with $v=1.3$~mm/s, but without spin-orbit coupling. We find that similar density modulations appear in the spin-orbit coupled lattice system for both translational directions of the lattice, e.g. in Fig.~\ref{SpinDynamics}(b) with $v=1.1$~mm/s, $\hbar\delta = 1.6~E_{\rm Ram}$, and Fig.~\ref{SpinDynamics}(c) with $v=-6.5$~mm/s, $\hbar\delta =  1.6~E_{\rm Ram}$.  The onset of the density modulations  in Fig.~\ref{SpinDynamics}(c), which is taken for a velocity near the $\underline{2}$ band edge, is accompanied with a momentum transfer of $2 \hbar k_{\rm lat}$ for a small but noticeable fraction of the atoms.  The transferred population, occupying $\hbar k_z \approx 0$, acquires nearly balanced spin composition.

\subsection{Single particle band structure}

While the effective dispersion relation presented in Fig.~2a allows the easy determination of the band edges, it can be seen from $\tilde{H}^M_{\rm sp}$ that a single band structure can be chosen by fixing $\delta_{\rm eff}$. To realize such a case, the lattice velocity and Raman detuning can always be adjusted together such that $\delta_{\rm eff}= \delta + 2m \gamma v /\hbar=$ constant. When varying $v$ under this constraint, the BEC effectively changes quasimomentum within a fixed dispersion relation. This is demonstrated in Fig.~\ref{SuppBandStructure} for $\hbar\delta_{\rm eff}$ fixed to $0, +3~E_{\rm Ram}$ and $-3~E_{\rm Ram}$.  For fixed $\delta_{\rm eff}$, the quasimomentum of the BEC is given by $m (\delta-\delta_{\rm eff}) /{2 k_{\rm Ram}}+k_{\rm min}$, where $k_{\rm min}$ is dependent on $\delta$.  This is showcased in the supplementary movie file for $\delta_{\rm eff}$ ranging from $11~E_{\rm Ram}$ to $-8~E_{\rm Ram}$ with the locations of the data taken in Fig.~3 overlaid.  Only values of positive $\delta$ can be probed within these fixed dispersion relations for the trapped BEC, which is prepared to the global minimum of the spin-orbit band structure. When $\delta$ changes sign, the quasimomentum of the BEC $k_{\rm min}$ changes sign as well. 

\bibliography{SOClattice}{}
 \end{document}